\documentclass[Print]{Style/isecure-v24}

\usepackage{lipsum}
\usepackage{algpseudocode,algorithm}
\usepackage[colorlinks, bookmarksnumbered, linkcolor=darkblue, citecolor=darkred, urlcolor=darkgreen]{hyperref}



\usepackage[numbers,sort&compress]{natbib}
\usepackage[all]{hypcap}
\usepackage{Style/picins}
\usepackage{lettrine}
\usepackage{wrapfig}
\usepackage{graphicx}
\usepackage{subcaption}
\usepackage{amsthm}
\usepackage{amssymb,amsmath,amsfonts,eqnarray,breqn}
\usepackage{fancyhdr}
\usepackage{ragged2e}
\usepackage{enumitem}   
\usepackage{refcount}
\usepackage[protrusion=true,expansion=true]{microtype}
\usepackage{float}
\usepackage[greek,english]{babel}
\usepackage[LGR,T1]{fontenc}
\usepackage{natbib}
\usepackage{pifont} 
\usepackage{array}  
\usepackage{wasysym} 
\usepackage{booktabs} 
\usepackage{adjustbox}
\usepackage{xcolor}

\DisableLigatures[f]{encoding=*,family=*}
\graphicspath{{Images/},{Style/}}
\DeclareGraphicsExtensions{.jpg,.pdf,.png}
\input{Style/Theorem-Styles.tex}
\articletype{Short Paper}

\usepackage{titlesec}
\titleformat{\paragraph}
{\normalfont\normalsize\bfseries}{\theparagraph}{1em}{}
\titlespacing*{\paragraph}
{0pt}{3.25ex plus 1ex minus .2ex}{1.5ex plus .2ex}

\journal{ISeCure}
\company{ISC}
\copyrightyear{2025}
\journalsite{http://www.isecure-journal.org}
\firstpage{1} 

\makeatletter
\begingroup \lccode`+=32 \lowercase
 {\endgroup \def\Url@ObeySp{\Url@Edit\Url@String{ }{+}}}
 \def\Url@space{\penalty\Url@sppen\ }
\makeatother

\usepackage[displaymath]{lineno}  
\newcommand*\patchAmsMathEnvironmentForLineno[1]{
  \expandafter\let\csname old#1\expandafter\endcsname\csname #1\endcsname
  \expandafter\let\csname oldend#1\expandafter\endcsname\csname end#1\endcsname
   \renewenvironment{#1}
     {\linenomath\csname old#1\endcsname}
     {\csname oldend#1\endcsname\endlinenomath}}
\newcommand*\patchBothAmsMathEnvironmentsForLineno[1]{
  \patchAmsMathEnvironmentForLineno{#1}
  \patchAmsMathEnvironmentForLineno{#1*}}
\AtBeginDocument{
\patchBothAmsMathEnvironmentsForLineno{equation}
\patchBothAmsMathEnvironmentsForLineno{align}
\patchBothAmsMathEnvironmentsForLineno{flalign}
\patchBothAmsMathEnvironmentsForLineno{alignat}
\patchBothAmsMathEnvironmentsForLineno{gather}
\patchBothAmsMathEnvironmentsForLineno{multline}
}

\begin{document}
\begin{frontmatter}


\def\NoDingTitle{Securing Deep Learning Hardware: A Survey of Side-Channel Vulnerabilities and Countermeasures}
\title{\NoDingTitle\textsuperscript{ }}


\author[a1]{Zahra Mohammadi},
\ead{zahramohammmadi@ut.ac.ir}
\author[a1]{Mona Hashemi}, and%
\ead{Hashemi.mona@ut.ac.ir}
\author[a1,CorAuth]{Siamak Mohammadi}
\ead{smohamadi@ut.ac.ir}

\address[a1]{School of Electrical and Computer Engineering, University of Tehran, Tehran, Iran}

\corauth[CorAuth]{Corresponding author.}

\begin{abstract}
As deep learning models are increasingly deployed in critical sectors such as healthcare, finance, and security, ensuring their protection against emerging threats has become crucial. Among these threats, side‑channel attacks (SCAs) represent a particular challenge since they can extract sensitive information such as model architectures, parameters, and even user inputs without requiring direct access to the model. By leveraging the physical and micro-architectural properties of the hardware, attackers can compromise systems. This survey begins by classifying leakage sources and attacker objectives, then analyzes representative studies that demonstrate practical side‑channel exploits against deep‑learning hardware. It also reviews existing defenses aimed at mitigating these vulnerabilities and concludes by outlining key open research challenges and potential future directions.
\end{abstract}


\makeatother

\end{frontmatter}

\addtolength{\parskip}{2mm}

\section{Introduction} \label{intro}

Deep learning models have become fundamental to a wide array of modern applications, including cloud computing, mobile platforms, Internet of Things (IoT) devices, and critical infrastructure. Their exceptional accuracy and performance have led to widespread deployment across both data centers and resource-constrained edge devices. However, the underlying hardware used to run these models such as Central Processing Units (CPUs), Graphics Processing Units (GPUs), Field-Programmable Gate Arrays (FPGAs), and custom accelerators can unintentionally expose sensitive information through physical and micro-architectural side channels.

Side-channel leakages, such as power consumption, memory access patterns, timing differences, and electromagnetic (EM) emissions, can be exploited by adversaries to extract critical model information. These include architecture details, model parameters, or even private user inputs~\cite{1,2,3}. Such attacks do not rely on software vulnerabilities but instead observe and analyze low-level physical behaviors of hardware during model execution.

This survey provides a comprehensive overview of the landscape of hardware side-channel vulnerabilities in deep learning models. We present a structured taxonomy of SCAs, analyze representative attack techniques, and review defensive strategies developed to mitigate these threats. By categorizing attacks based on leakage sources, attacker capabilities, and objectives, this paper aims to shed light on the emerging security challenges and motivate the development of more resilient machine learning hardware systems.

\subsection{Deep Learning Model as Intellectual Property (IP)}

Modern deep learning models represent far more than simple lines of code—they are valuable forms of IP, developed through extended efforts involving expert engineering, costly training procedures, and often the use of proprietary datasets. Constructing a competitive model typically requires large-scale data collection and annotation, careful design of neural network architectures, and thorough hyperparameter tuning. For instance, training large-scale models such as Generative Pre-trained Transformers like GPT-3 and GPT-4 is estimated to cost over \$100 million, with expenses often offset through monetization strategies like paid services or Application Programming Interfaces (APIs) \cite{26}.

Beyond financial investment, these models encapsulate strategic knowledge that can serve as a major competitive advantage. Many are fine-tuned using proprietary data—ranging from user behavior to medical records—which enables them to outperform publicly available or open-source alternatives. If an attacker successfully steals such a model, they can reproduce its functionality without incurring the original development costs. This constitutes IP theft and poses a serious threat to the developer’s market position \cite{4}.

Furthermore, deep learning models can unintentionally retain portions of their training data, especially when overfitting or fine-tuning occurs on small, sensitive datasets. This raises serious privacy concerns in domains such as healthcare, finance, and biometrics, where data leakage could lead to regulatory violations or facilitate reconstruction attacks such as membership inference and model inversion~\cite{4}.

These risks are not merely theoretical. In a recent controversial case, it was suspected that DeepSeek had copied models developed by OpenAI without explicit permission \cite{25}. While no definitive evidence of model theft has been confirmed, the case raised serious concerns about the potential for unauthorized use of proprietary artificial intelligence (AI) systems. Such incidents highlight the growing importance of protecting deep learning models as strategic assets.

\subsection{Threat Landscape for Deep Learning Hardware}

The widespread deployment of deep learning across cloud platforms, edge devices, and embedded systems has significantly expanded the risk of SCAs. The hardware used to execute these models has historically been designed with a primary focus on performance, not security or side-channel resilience. As a result, these devices may unintentionally expose sensitive information through hardware-level leakage. Adversaries can exploit observable signals such as power consumption, EM radiation, cache behavior, and timing differences to extract confidential details, including model parameters and user inputs ~\cite{1,2,3,28,29,30}.

One of the most prominent vulnerabilities emerges in multi-tenant cloud environments, where computational resources like CPUs and GPUs are shared among numerous users, some of whom may not be trusted. This setup is typical in commercial platforms such as Amazon Web Services (AWS) and Microsoft Azure \cite{5}. In such settings, attackers can apply techniques such as timing analysis, context-switch monitoring, and cache probing to recover model parameters or reconstruct inputs, all without breaching standard hardware isolation protocols ~\cite{1,5,6,23,33,34}.

Another critical concern lies in edge deployments, including applications such as smart cameras, drones, autonomous vehicles, and IoT sensors. These devices are often physically accessible and therefore particularly susceptible to power analysis and EM SCAs. Attackers can connect probes to power lines or observe EM fields to infer internal model information or reconstruct private inputs such as images. The physical nature of these attacks makes them difficult to detect and mitigate—especially in devices with limited processing capabilities, minimal tamper resistance, or cost-sensitive designs ~\cite{2,4,12,31,32}.

Taken together, these scenarios reveal a growing and diverse attack surface for deep learning hardware. From EM attacks on consumer drones to cache-based inference in large-scale data centers, adversaries have multiple vectors through which they can bypass software-layer protections and compromise the confidentiality of deployed models.

\subsection{Research Motivation}

While SCAs have been extensively studied in the context of cryptographic hardware, their implications for deep learning systems remain comparatively underexplored. The hardware platforms that execute these models pose unique challenges that are not fully addressed by traditional security models \cite{35}.
One critical distinction lies in the nature and scale of the underlying secrets. Unlike cryptographic keys, which have short and fixed-length representations, deep neural networks may contain millions or even billions of floating-point parameters. The leakage of these parameters can differ significantly across layers, depending on both model architecture and input data, which complicates detection and mitigation \cite{35}.

Another difficulty stems from the dynamic and irregular execution patterns of deep learning workloads. Variable input dimensions, mixed-precision arithmetic, compiler optimizations, and hardware-level scheduling all introduce significant noise and unpredictability into side-channel traces. These factors increase the complexity of constructing effective attacks and designing robust defenses, particularly in multi-tenant or real-time execution environments.

At the same time, hardware innovation is advancing faster than security evaluation can keep pace. New classes of AI hardware such as tensor cores, neuromorphic chips, and photonic processors are being introduced into the market before their side-channel vulnerabilities are fully understood.

Beyond technical concerns, these challenges also pose serious economic risks. As Machine Learning as a Service (MLaaS) platforms and embedded AI solutions generate billions of dollars in revenue, a successful SCA could result in IP theft, privacy breaches, and major financial losses. Yet, most existing work focuses on algorithmic threats such as adversarial examples or cryptographic protections, while overlooking hardware-level leakage channels.

This survey aims to bridge that gap by offering:

\renewcommand{\labelitemi}{$\bullet$}
\begin{itemize}
	\item A unified taxonomy of hardware SCAs on deep learning models, highlighting leakage sources, attacker capabilities, and attack goals.
	\item A review of representative attacks reported in the literature.
	\item An overview of existing defense mechanisms and runtime mitigation strategies.
	\item An analysis of current research gaps and future directions, including potential threats to transformer-based models and other emerging AI frameworks.
\end{itemize}

\subsection{Literature Search Methodology}
To provide a comprehensive overview of hardware side-channel vulnerabilities in deep learning models, we employed a systematic literature search strategy. Initially, several key survey papers in the domain were reviewed to understand its breadth; this field encompasses diverse hardware platforms (such as CPUs, GPUs, FPGAs, and custom accelerators) and various leakage sources that are exploited to extract three primary categories of model information (architecture, parameters, and inputs), each of which holds significant value for adversaries. Based on these insights, the taxonomy presented in Figure~\ref{fig:sca-taxonomy} was developed. Subsequently, priority was given to leakage sources; for each source, papers with high citation counts or foundational works were first identified and studied, followed by more recent (up-to-date) contributions, and finally, studies focusing on different hardware targets (such as cache-based side-channel attacks on both CPUs and GPUs) were selected to ensure diverse coverage.
The search primarily focused on papers published in the last five years (2020–2025), as this is an emerging field with most relevant research falling within this timeframe. Inclusion criteria encompassed papers that specifically addressed side-channel attacks against hardware platforms used in deep learning models and provided empirical evidence or theoretical analyses of attack methods, vulnerabilities, or countermeasures. Exclusion criteria eliminated papers primarily centered on non-hardware attacks or information extraction from non-deep learning models. This approach, while comprehensive, concentrates on approximately 30–40 representative studies to maintain analytical depth and avoid redundancy.

The rest of this paper is organized as follows: Section~\ref{background} introduces deep learning hardware platforms, identifies major leakage sources, and presents a taxonomy of SCAs. Section~\ref{three} reviews several representative attacks categorized by their leakage type. Section~\ref{four} surveys defense mechanisms and detection techniques. Section~\ref{five} outlines open research challenges and future directions. Finally, Section~\ref{six} concludes the paper.

\section{Background }
\label{background}

Before exploring specific SCA techniques and defense mechanisms, this section provides the necessary background and introduces a structured taxonomy. SCAs exploit unintentional information leakage from physical or micro-architectural behaviors of hardware such as timing variations, power consumption, EM emissions, or cache activity to extract sensitive data. Unlike conventional software-based attacks that target code-level vulnerabilities, SCAs operate by passively or actively monitoring low-level hardware behavior during model execution.

This section begins by surveying the landscape of deep learning hardware platforms and identifying prevalent sources of information leakage. It then outlines typical attacker capabilities and threat models, followed by a classification of attack objectives and the ways in which extracted information may be exploited. To provide a practical understanding, representative attack primitives such as power analysis and cache-based methods are also introduced.

\begin{figure*}[htbp]
	\centering
	\includegraphics[width=16cm, height= 7cm]{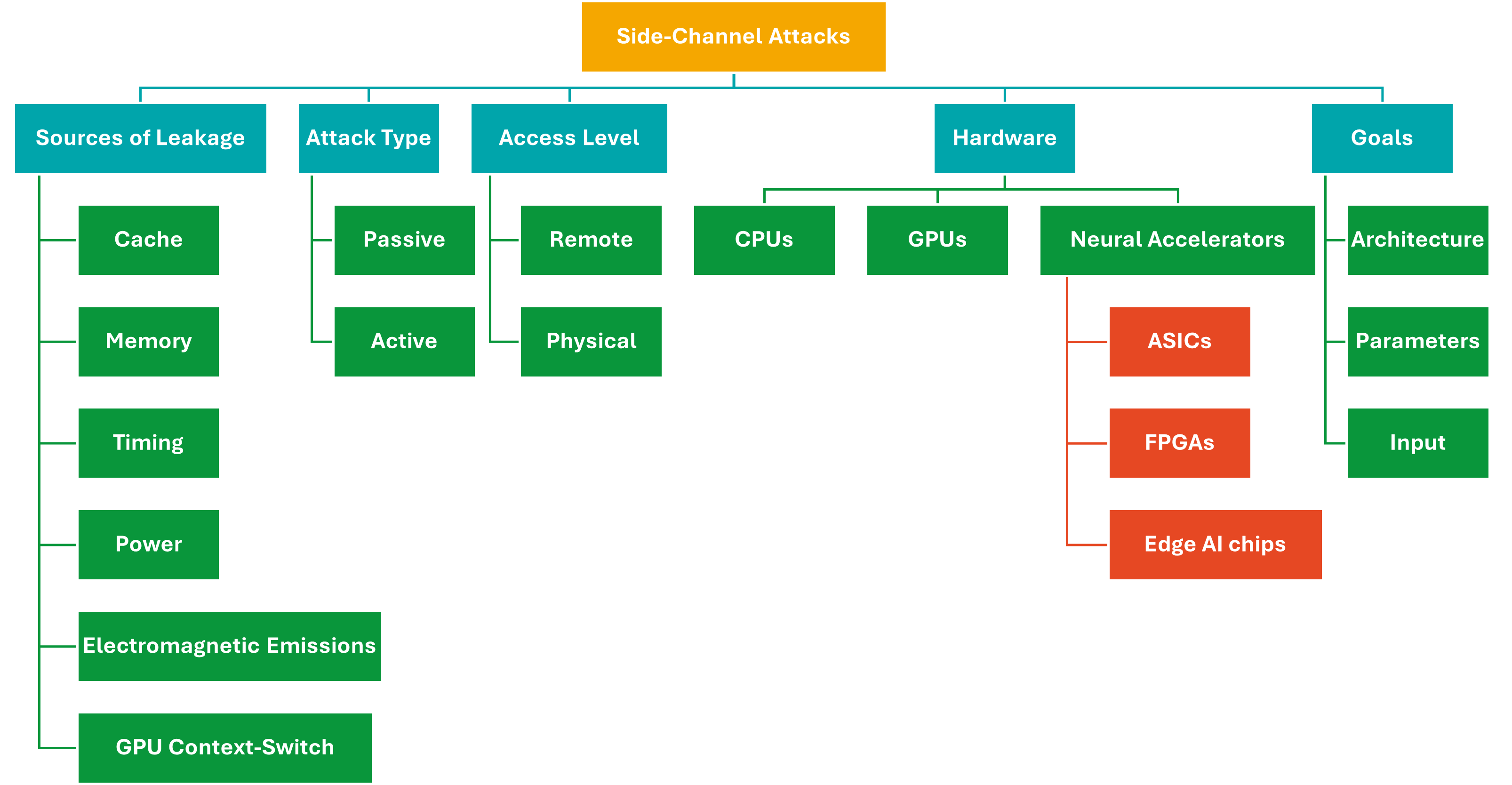}
\caption{
	Overall taxonomy of hardware SCAs on deep learning models
}
	\label{fig:sca-taxonomy}
\end{figure*}

Figure~\ref{fig:sca-taxonomy} presents a comprehensive taxonomy of hardware SCAs on deep learning models, categorized according to key dimensions such as leakage sources, attack methodology, access level, target hardware, and attacker objectives. While this taxonomy draws inspiration from general SCA classifications in prior surveys, such as the one proposed in \cite{2}, it is specifically tailored to deep learning systems. For instance, unlike general SCA taxonomies that emphasize broad categories like logical vs. physical exploited properties or profiling phases across various cryptographic implementations, our taxonomy highlights deep learning specific elements, including hardware platforms optimized for neural networks (e.g., GPUs, FPGAs, and edge AI chips) and attacker goals focused on extracting model architecture, parameters, or inputs. This adaptation addresses the unique vulnerabilities in machine learning hardware deployments.

\subsection{Deep Learning Hardware}

To meet the growing computational demands of deep learning, specialized hardware platforms are widely adopted. However, each of these platforms introduces distinct vulnerabilities and potential side-channel leakage vectors that have been explored in various studies.

$\bullet$\textbf{CPUs} are commonly used for neural network inference, particularly in scenarios where flexibility, energy efficiency, or cost-effectiveness is prioritized over raw performance. Due to their extensive use of shared memory resources, especially the Last-Level Cache (LLC), CPUs are highly susceptible to cache-based SCAs. For instance, studies such as \textit{DeepCache} and \textit{Cache Telepathy} have demonstrated that techniques like \textit{Flush+Reload} can accurately recover the architecture and parameters of convolutional neural networks (CNNs) running on CPU-based platforms~\cite{5,7}. Additionally, the work in \cite{14} demonstrates a stealthy inference attack using cache-based side-channel on CPUs to extract the model outputs, while the article \textit{DeepTheft} \cite{31} exploits power side channels to infer model architectures on CPUs.

$\bullet$\textbf{GPUs}, with their massively parallel architecture optimized for high-throughput numerical operations, are ideal for both training and batch inference of deep learning models. Despite their advantages, GPUs exhibit significant side-channel vulnerabilities due to parallel execution, predictable scheduling patterns, and distinctive power consumption signatures. For example, the \textit{Leaky DNN} attack exploited timing variations during context switching to extract detailed architectural information from models running on shared GPU infrastructure~\cite{6}. Similarly, the \textit{BarraCUDA} attack leveraged EM emissions from GPU operations to recover the weights of convolutional layers with high precision~\cite{8}. 
Additionally, the work in \cite{36} demonstrates the exploitation of side-channel information from GPUs to extract optimized Deep Neural Network (DNN) architectures, using techniques like timing variations during inference. Similarly, the article \textit{Spy in the GPU-box} \cite{37} highlights a covert channel attack across multi-GPU systems, where cache contention across GPUs was used to infer sensitive data, including model information from a remote GPU.

$\bullet$\textbf{FPGAs} are reconfigurable and flexible platforms that support rapid prototyping and the deployment of custom accelerators for deep learning inference tasks. Likewise,custom accelerators—such as tensor processing units and neuromorphic chips—use fixed-function hardware to optimize computational efficiency and energy consumption. However, both FPGAs and custom accelerators may expose side-channel vulnerabilities through predictable patterns in power usage and EM emissions during intensive computation. For instance, researchers \cite{9} demonstrated that memory access patterns in CNN accelerators can be exploited to reverse-engineer the architecture and parameters of encrypted models. Moreover,  another work \cite{10} showed that FPGA-based CNN accelerators could leak input image information through power side-channels, enabling attackers to reconstruct sensitive inputs without requiring knowledge of the model's internal parameters. These examples underscore the critical need for dedicated defense mechanisms tailored to FPGA-based and custom deep learning hardware implementations. 
Additionally, \cite{39} presents a novel remote power attack on Cloud FPGAs that allows attackers to steal sensitive input data of DNN models through power side-channels. \cite{40} introduces a method to recover inputs from neural network accelerators on FPGAs using generative CNNs, highlighting the effectiveness of power-based side-channel attacks for input recovery. \cite{41} leverages a three-dimensional power surface to recover input images, showing the advanced capabilities of power analysis in FPGA-based DNN accelerators. Finally, \cite{42} proposes the \textit{NNLeak} attack, which successfully extracts full DNN models by combining power analysis with multi-stage correlation techniques, demonstrating the feasibility of these attacks on AI-oriented accelerators.

\subsection{Sources of Side-Channel Leakage}

SCAs exploit indirect leakages that result from hardware execution to extract sensitive information. These leakages stem from the physical or micro-architectural behavior of hardware and may vary across different platforms. This subsection categorizes the primary leakage sources exploited in attacks targeting deep learning hardware.

$\bullet$\textbf{Cache Leakage} occurs when attackers exploit access patterns and timing behaviors within the shared cache hierarchy—particularly in CPUs and GPUs. These attacks determine whether specific memory blocks remain cached after victim execution, thereby revealing neural layer activations or tensor usage. Techniques such as \textit{Flush+Reload} and \textit{Prime+Probe} have been widely used to recover model architecture information. For instance, the \textit{DeepCache} and \textit{Cache Telepathy} and other studies demonstrate how CNN structures can be reconstructed by probing shared libraries (e.g., BLAS) during inference on shared CPU platforms~\cite{5,7,21,22,23}.

$\bullet$\textbf{Memory Leakage} results from observing the behavior of the physical memory subsystem, including buffer activity. Attacks in this category trace memory access patterns to extract neural network parameters. For example,~\cite{9} shows how CNN accelerators implemented on custom hardware can be reverse-engineered by analyzing memory access traces.

$\bullet$\textbf{Timing Leakage} is based on measuring inference execution time under varying input conditions. Differences in latency may stem from layer types (e.g., convolution vs. pooling), branching conditions, or optimization paths. Attackers can leverage these timing variations to infer the model’s depth, structural features, or even properties of the input data \cite{23}. Additionally, \cite{43} demonstrates a timing attack on sparsity-aware neural networks, where timing side-channel information from optimizations like skipping sparse multiplications is used to predict the model’s classification outcomes, revealing sensitive architectural and input data details.

$\bullet$\textbf{Power Leakage} exploits variations in power consumption during model execution. By monitoring these changes using power sensors or side-channel monitors, attackers can correlate specific patterns with operations such as multiply–accumulate (MAC). Simple power analysis (SPA) can reveal layer boundaries, while differential power analysis (DPA) can extract parameters like weights or input bits. The study in~\cite{4} demonstrates how DPA combined with power correlation techniques can recover convolutional weights.

$\bullet$\textbf{EM Leakage} results from unintended radio-frequency emissions during hardware operation. These signals are strongly correlated with switching activity in circuits and can be captured using near-field probes. EM SCAs have proven especially effective in embedded architectures such as GPUs and low-power accelerators. For example, the \textit{BarraCUDA} attack~\cite{8} utilized EM leakage from NVIDIA GPUs to extract convolutional weights, successfully reconstructing model parameters with high accuracy.

$\bullet$\textbf{Context Switch Leakage} targets multi-tenant GPU systems, where multiple workloads share the GPU through time-sliced execution. When the GPU switches from one process to another, residual register states and timing artifacts may unintentionally leak sensitive information. The \textit{Leaky DNN} attack demonstrates this by recovering architectural details and layer dimensions through delays observed during context switches in shared GPU environments~\cite{6}.

\subsection{Adversary Capabilities and Access Methods}

This subsection categorizes threat models based on two primary dimensions: the adversary’s mode of interaction with the system (passive vs. active) and their method of access (physical vs. remote). These classifications help determine which types of attacks are realistic in different deployment environments, such as cloud servers or edge devices.

$\bullet$\textbf{Passive vs. Active Attacks:}
In passive SCAs, the adversary silently observes unintentional leakages—such as power consumption, cache behavior, timing differences, or EM emissions—to extract sensitive information. These attacks are typically stealthy and difficult to detect. In contrast, active SCAs deliberately interfere with the system to induce or amplify information leakage. Techniques may include fault injection, clock/voltage manipulation, or thermal tampering. Active attacks can expose system states that would otherwise be inaccessible but often leave detectable traces, making them easier to identify~\cite{1,2}.

$\bullet$\textbf{Physical vs. Remote Access:}
Physical attacks assume that the adversary has direct access to the hardware device—such as placing a probe on a smart camera or IoT sensor, or measuring emissions from a chip in a laboratory environment. This level of access enables high-resolution monitoring of EM signals or power traces. On the other hand, remote attacks are executed entirely through software interfaces or virtual environments without requiring physical contact. These attacks are common in shared cloud infrastructures, where adversaries can exploit cache-based side channels or delays introduced by GPU context switching to extract sensitive information~\cite{1,2}.

\subsection{Attack Objectives and Potential Post-Extraction Abuses}

In hardware SCAs targeting deep learning models, adversaries typically aim to extract valuable internal information from neural networks—such as the model’s architecture, parameters, or input data. Each of these targets holds significant value for attackers and can enable a range of subsequent malicious activities.

\textbf{Attack Objectives:}

$\bullet$ \textbf{Architecture:}
Architecture extraction involves revealing the internal structure of a neural network, including the number and types of layers (e.g., convolutional, fully connected, pooling), the number of neurons or filters in each layer, activation functions, and inter-layer connections. Knowing the architecture effectively gives the adversary the model’s blueprint, which facilitates further attacks such as model inversion~\cite{2,4}.

$\bullet$ \textbf{Parameters:}
Parameters refer to the numerical weights and biases learned during training. These values directly influence how inputs are transformed at each layer and ultimately determine the model’s outputs. Recovering these parameters allows the adversary to replicate the model’s behavior completely~\cite{2,4}.

$\bullet$ \textbf{Inputs:}
Inputs include the actual data used during training or inference, such as sensitive images, textual data, or private records (e.g., medical scans or financial transactions). Extracting input data poses serious privacy risks and may lead to the exposure of confidential or regulated information~\cite{2,4}.

\textbf{Potential Post-Extraction Abuses:}

$\bullet$ \textbf{Evasion:}
In evasion attacks, adversaries craft adversarial examples—inputs that appear benign to humans but cause the model to produce incorrect results. For instance, a minor pixel change in a cat image could cause it to be misclassified as a dog. These attacks undermine the reliability and robustness of deployed models \cite{27}.

$\bullet$ \textbf{Poisoning:}
Poisoning involves inserting malicious or misleading data into the training set to corrupt the learning process. This can cause the model to behave incorrectly or embed hidden backdoors. For example, attackers may inject modified stop sign images that later cause misclassification in real-world settings. This compromises the model’s integrity and may lead to unsafe decisions \cite{27}.

$\bullet$ \textbf{Membership Inference:}
Membership inference seeks to determine whether a particular data point was used in training. This can severely breach privacy, especially in domains like healthcare or finance. For example, an attacker may infer whether someone’s medical record was part of the training dataset, raising compliance and ethical concerns \cite{27}.

$\bullet$ \textbf{Model Inversion:}
Model inversion attacks aim to reconstruct input data using model outputs or internal behaviors. For instance, an attacker with access to the output of a facial recognition model might regenerate a recognizable image of a face. This poses a serious privacy risk for models trained on sensitive or biometric data \cite{27}.

$\bullet$ \textbf{Model Stealing:}
In model stealing attacks, adversaries attempt to duplicate the neural network either by extracting its architecture and parameters or by creating an approximate replica based on observed outputs. Stolen models allow attackers to offer similar services without training costs, undermining the original developer’s competitive edge and causing direct financial harm \cite{27}.

\newcolumntype{C}[1]{>{\centering\arraybackslash}p{#1}}

\begin{table}[hb]
	\centering
	\begin{adjustbox}{max width=1\linewidth} 
		\begin{tabular}{@{}l|C{1.2cm}C{1.4cm}C{1.8cm}C{1.4cm}C{1.4cm}@{}}
			\toprule
			\textbf{Objective}&\textbf{Evasion}&\textbf{Poisoning}&\textbf{Membership Inference}&
			\textbf{Model Inversion}&\textbf{Model Stealing}\\
			\midrule
			Arch.   & $\CIRCLE$ & $\LEFTcircle$ & $\Circle$ & $\Circle$ & $\CIRCLE$\\
			Params. & $\CIRCLE$ & $\CIRCLE$     & $\CIRCLE$ & $\CIRCLE$ & $\CIRCLE$\\
			Input   & $\LEFTcircle$ & $\LEFTcircle$ & $\Circle$ & $\Circle$ & $\LEFTcircle$\\
			\bottomrule
		\end{tabular}
	\end{adjustbox}
	\caption{Effect of each attack objective on corresponding post‑extraction abuses~\cite{4}.}
	\label{tab:attack-objectives}
\end{table}

Table~\ref{tab:attack-objectives} illustrates the relationship between different attack objectives and their potential post-extraction abuses. It visualizes how the extraction of architecture, parameters, or input data can enable various adversarial activities such as evasion, poisoning, membership inference, model inversion, and model stealing. The level of impact is represented using filled, half-filled, and empty circles to denote strong, moderate, or limited relevance, respectively.

\subsection{Representative SCAs}

This section presents a detailed overview of SCA techniques frequently reported in the literature. Specifically, we focus on two prominent categories of leakage exploited in these attacks: power and EM analysis, and cache-based analysis. These approaches have repeatedly demonstrated their effectiveness in compromising deep learning hardware.

\subsubsection{Power and EM Analysis Attacks}

Power consumption and EM emissions are among the most commonly exploited physical side channels. When deep learning models are executed on hardware, switching activity within logic units and memory components produces distinct power and EM patterns. Among numerous analysis techniques, SPA and DPA have emerged as two of the most effective and widely adopted methods. This section describes how each technique operates and how they are employed to extract model-level details such as architecture and parameters in the deep learning context.

$\bullet$	\textbf{SPA} involves direct visual inspection of a single power or EM trace to identify distinguishable operational patterns. It takes advantage of the fact that different computational operations (e.g., convolution, fully connected layers, activation functions) typically produce unique power signatures. By visually examining recorded traces, an attacker can infer specific operations executed by a neural network without the need for complex statistical techniques.
	For example, SPA can reveal the boundaries between layers in a neural network, such as identifying the end of convolution operations and the start of activation functions (e.g., ReLU, Softmax), simply by observing fluctuations in power during inference. This enables adversaries to reconstruct a deep learning model using minimal effort and without requiring multiple measurements.
	
	The following steps illustrate a typical SPA attack aimed at extracting a model’s architecture:
	
	\begin{enumerate}
		\item \textit{Trace Collection:} The attacker passively captures a power or EM trace during model inference on target hardware, such as an FPGA.
		\item \textit{Visual Inspection:} The attacker visually inspects the collected trace to identify unique signal patterns or spikes. For instance, convolutional layers often show repetitive, easily distinguishable patterns due to intensive MAC operations.
		\item \textit{Layer Identification:} Based on transitions between distinct patterns (e.g., from repetitive convolution to simpler activation), the attacker identifies layer types and boundaries.
		\item \textit{Model Reconstruction:} Once operations and their sequence are identified, the attacker reconstructs the network’s architecture—inferring the number of layers, layer types (e.g., convolution, activation, pooling), and their approximate order.
	\end{enumerate}
	
	This step-by-step approach highlights SPA’s simplicity and effectiveness, showing how adversaries can uncover critical architectural information using basic visual inspection techniques.
	
$\bullet$	\textbf{DPA} extends SPA by statistically correlating multiple power or EM traces with hypothesized numerical values—typically weights, biases, or inputs. Instead of analyzing a single trace, the attacker collects hundreds to millions of traces from repeated inferences on controlled, known inputs and applies statistical tests (e.g., Pearson correlation) to determine which guessed values best explain the measured leakage. Studies show that with sufficient traces and precise alignment, DPA can recover highly accurate weight values from deep learning models.
	A typical DPA attack targeting model weights proceeds as follows:
	
	\begin{enumerate}
		\item \textit{Trace Collection:} The attacker feeds a set of known inputs into the target hardware and records a power trace for each inference.
		\item \textit{Segmentation and Alignment:} Each trace is segmented so that each portion aligns with a distinct MAC operation or layer.
		\item \textit{Hypothesis Generation:} For a candidate weight value and known input, the attacker computes a predicted leakage signal.
		\item \textit{Statistical Correlation:} Correlation between each candidate’s leakage vector and real trace samples is computed; the weight with the highest correlation is assumed correct.
		\item \textit{Iterative Recovery:} The process continues for subsequent weights, using previously recovered ones, until a full layer is reconstructed.
		\item \textit{Validation and Refinement:} The model is re-executed with new inputs to verify if the recovered parameters yield acceptable outputs; otherwise, more traces are collected or alignment is adjusted.
	\end{enumerate}
	
	Because DPA yields precise numerical parameters, it enables direct model replication and facilitates further attacks such as membership inference or model inversion. Its primary drawbacks are the heavy trace collection overhead and the need for fine-grained synchronization. However, when these challenges are addressed, DPA stands out as one of the most powerful and threatening hardware side-channel techniques against deep learning models.

\subsubsection{Cache-Based Attacks}

\begin{figure}[htbp]
		\centering
		\includegraphics[width=1\linewidth]{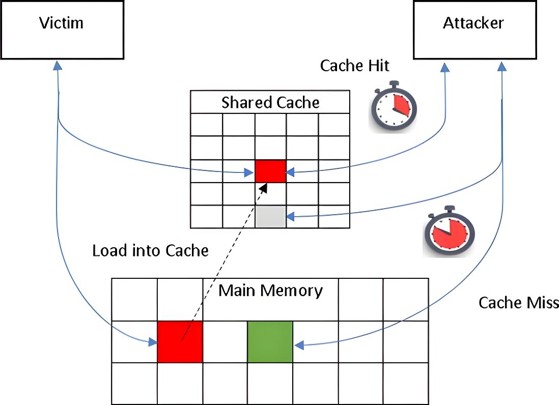}
		\caption{Generic overview of cache side‑channel attacks \cite{24}.}
		\label{fig:Cache}
	\end{figure}

Many hardware platforms use multi-level cache architectures. These caches often leave measurable timing footprints when lines are flushed, loaded, or reloaded. Cache-based SCAs exploit such timing leakages to infer a victim's memory access patterns without requiring direct access to their memory. This subsection outlines three widely adopted techniques used to exploit caches, each accompanied by a step-by-step explanation to illustrate its operation in practice. As illustrated in Figure~\ref{fig:Cache}, a typical cache‑based side‑channel scenario involves a co‑located attacker core and a victim core that contend for a shared LLC. By priming, flushing, or probing specific cache sets, the attacker distinguishes cache hits from misses and thereby infers the victim’s memory‑access footprint.

	
	$\bullet$\textbf{Flush + Reload:}This technique applies when the attacker and victim share memory pages (e.g., shared libraries). It uses the \texttt{clflush} instruction to flush cache lines and measures reload time to determine if the victim accessed them \cite{44}.The main steps of this attack are:
	\begin{enumerate}
		\item \textit{Shared memory mapping:} The attacker maps a memory region shared with the deep learning model.
		\item \textit{Flush:} Specific cache lines are flushed using \texttt{clflush}.
		\item \textit{Victim execution:} The model may reload those lines.
		\item \textit{Reload and timing:} The attacker reloads the lines and measures access latency.
		\item \textit{Analysis:}
		\begin{itemize}
			\item Low latency $\Rightarrow$ victim used the line.
			\item High latency $\Rightarrow$ line was unused.
		\end{itemize}
		\item \textit{Repeat:} To reconstruct the model’s memory access pattern.
	\end{enumerate}
	
	$\bullet$ \textbf{Prime + Probe:}Unlike Flush + Reload, this technique does not require shared memory. It fills a cache set (prime), waits for the victim to execute, then re-accesses the same set (probe) to measure eviction \cite{5}. General steps of this attack are:
	\begin{enumerate}
		\item \textit{Prime:} The attacker fills cache sets using their own data.
		\item \textit{Victim execution:} May evict the attacker’s data.
		\item \textit{Probe:} The attacker re-accesses the same addresses.
		\item \textit{Analysis:}
		\begin{itemize}
			\item High latency $\Rightarrow$ victim accessed that cache set.
			\item Low latency $\Rightarrow$ victim did not access it.
		\end{itemize}
		\item \textit{Repeat:} To build a temporal access map of the model.
	\end{enumerate}
	
	

\section{Hardware SCAs on Deep Learning models Case Studies}
\label{three}

This section provides a comprehensive review of recent Hardware SCAs targeting deep learning models. We analyze a selection of representative attacks published in the literature and categorize them based on their underlying leakage sources: power and EM emissions, cache and timing behaviors, memory access patterns, and context-switching mechanisms in GPUs. Each subsection presents a focused analysis of impactful studies, discussing their threat models, attack goals, methodologies, and key outcomes. The reviewed works highlight how various side-channel techniques can effectively extract sensitive information from deep learning systems and underscore the pressing need for robust countermeasures.
\subsection{Power and EM-Based Attacks}

Power and EM SCAs exploit physical signals emitted by hardware during computation to extract sensitive information about deep learning models. These signals, directly stemming from electrical activity during neural network execution, can reveal detailed insights about a model's architecture, parameters, or even its input data. The accuracy of such attacks largely depends on factors such as measurement precision, alignment of recorded traces, and noise reduction techniques. Despite the deployment of both software- and hardware-level protections, modern deep learning platforms—including GPUs and FPGAs—remain vulnerable to these forms of side-channel leakage~\cite{4,8,10,11}. In what follows, we review two representative studies that demonstrate the feasibility of recovering neural network internals via power and EM side channels.

	 \subsubsection{SoK: Extracting Neural Network Secrets via Physical Side Channels}
	
	An attack ~\cite{4} that focuses on side-channel leakage arising from power and EM emissions with the goal of recovering confidential information from deep neural networks, including model architecture, trained parameters (weights and biases), and sensitive inputs. The attacks are evaluated on a commercial FPGA-based accelerator from Xilinx.
	
	\textbf{Attack Methodology:}

$\bullet$ \textbf{Architecture Extraction (via SPA):}
		
		\begin{enumerate}
			\item \textit{Trace Collection:} Power and EM signals are recorded during the neural network’s execution on known inputs.
			\item \textit{Layer Identification:} Using SPA, distinct patterns for each layer type are identified:
			\begin{itemize}
				\item Convolution: repetitive patterns with characteristic peaks.
				\item Activation functions such as ReLU (sharp spikes), Softmax (smooth curves), Tanh, and Sigmoid (distinct power profiles).
			\end{itemize}
			\item \textit{Layer Counting:} For simpler models, the number of layers can be deduced from trace repetitions. This becomes harder on parallel hardware like FPGAs.
			\item \textit{Hyperparameter Estimation:} Execution time patterns and model type allow inference of kernel sizes, strides, and neuron counts.
		\end{enumerate}
		
$\bullet$\textbf{Parameter Recovery (via DPA):}
		
		\begin{enumerate}
			\item \textit{Assumptions:} The attacker knows the model architecture and can supply chosen inputs.
			\item \textit{Weight Guessing:} Candidate values (e.g., 16-bit precision) are iteratively tested.
			\item \textit{Power Simulation:} Theoretical traces are generated based on weight guesses.
			\item \textit{Correlation Analysis:} Pearson correlation between simulated and real traces identifies correct weights.
			\item \textit{Error Propagation Issue:} Early weight errors distort deeper layers, making high accuracy critical. Even minor noise can cause significant accuracy degradation.
		\end{enumerate}
		
	$\bullet$ \textbf{Input Recovery (via DPA and Signal Analysis):}
		
		\begin{enumerate}
			\item \textit{DPA-Based:}
			\begin{itemize}
				\item Requires known architecture and parameters.
				\item Input values (e.g., pixel intensities) are guessed, simulated, and matched to measured traces.
			\end{itemize}
			\item \textit{Model-Free Classification:}
			\begin{itemize}
				\item EM traces from known input classes are used to train a classifier.
				\item New inputs are classified based solely on trace features; for MNIST, accuracy reaches approximately 88
			\end{itemize}
		\end{enumerate}

\begin{figure}[ht]
		\centering
		\includegraphics[width=\linewidth]{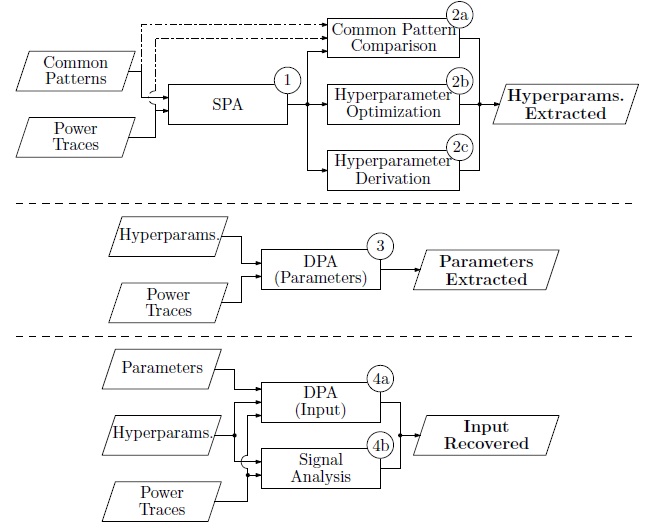}
		\caption{%
			Overview of the multi-stage SCA pipeline~\cite{4}.}
		\label{fig:SokPipeline}
	\end{figure}

	As summarised in Figure~\ref{fig:SokPipeline}, the attack unfolds in
	three main stages: architecture extraction, parameter recovery, and
	input reconstruction.
	This attack is especially effective against quantized and binary neural networks due to the smaller search space for weight values.
\subsubsection{BarraCUDA: Electromagnetic Leakage from GPUs Reveals Neural Network Parameters}
	
	BarraCUDA~\cite{8} presents a SCA based on EM leakage targeting neural networks deployed on NVIDIA Jetson Nano GPUs. The primary goal is to extract sensitive model parameters, including weights and biases.
	
	\textbf{Attack Methodology:}
	
	This attack uses Correlation EM Analysis (CEMA), a variant of DPA, to incrementally recover convolutional layer parameters:
	
	\begin{enumerate}
		\item \textit{Threat Model:} The attacker has physical access to the device, knows the model architecture, and can feed chosen inputs.
		
		\item \textit{Measurement Setup:}
		\begin{itemize}
			\item An EM probe is precisely placed near the GPU core.
			\item Around 15 million traces are recorded over 10 days using repeated inferences with known inputs.
		\end{itemize}
		
		\item \textit{Leakage Identification (via TVLA):}
		\begin{itemize}
			\item Statistical Test Vector Leakage Assessment (TVLA) pinpoints locations in the EM traces where weights or biases influence the signal.
		\end{itemize}
		
		\item \textit{Weight and Bias Extraction (via CEMA):}
		\begin{itemize}
			\item Inputs are known; possible weight values are hypothesized.
			\item For each guess, expected EM traces are simulated.
			\item Pearson correlation between simulated and real traces reveals the correct parameters.
		\end{itemize}
	\end{enumerate}
	
	The results reveal that even commercial, closed-source GPU platforms (like NVIDIA’s TensorRT) are susceptible to advanced SCAs based on EM leakage.

\subsection{Cache and Timing-Based Attacks}

Cache based SCAs exploit observable timing variations and memory access patterns during the execution of deep learning models on CPUs and GPUs. By carefully analyzing the behavior of shared hardware resources—especially the cache hierarchy in CPUs—these attacks can infer sensitive information about the model, including its architecture and layer configurations. Such attacks are particularly effective in shared computing environments, such as cloud infrastructures, where attackers often share physical resources with victims~\cite{5,7,21,22,23}. Below, we review two representative works that successfully demonstrate the extraction of neural network details via cache and timing behaviors.

	\subsubsection{Cache Telepathy: Leveraging Shared Resource Attacks to Extract Deep Neural Network Architectures}
	
The study in~\cite{5} focuses on recovering the architecture of deep neural networks in shared environments, such as cloud services and MLaaS platforms. The attacker runs on the same CPU as the victim and shares cache resources, enabling effective exploitation of cache side channels. The attack, known as Cache Telepathy, combines two classical cache side-channel techniques, Prime+Probe and Flush+Reload, to gradually extract structural details of the target network. It is demonstrated on popular architectures like VGG-16 and ResNet-50, using linear algebra libraries such as OpenBLAS and Intel MKL.
	
	\textbf{Attack Methodology:}
	\begin{enumerate}
		
		
		\item \textit{Inferring Matrix Multiplication Dimensions}:
		\begin{itemize}
			\item GEMM operations underpin the execution of most neural layers.
			\item By monitoring function calls like \texttt{itcopy}, \texttt{oncopy}, and \texttt{kernel}, the attacker can infer matrix dimensions involved in each layer.
			\item Example: if matrices $A$ ($m \times k$) and $B$ ($k \times n$) produce $C$ ($m \times n$), repeated access patterns reveal $m$, $n$, and $k$.
		\end{itemize}
		
		\item \textit{Mapping Dimensions to Network Parameters}:
		\begin{itemize}
			\item Fully-connected layers: matrix dimensions reveal neuron counts.
			\item Convolutional layers: dimensions are mapped to kernel size, padding, and stride using known relationships.
			\item Example: A recovered filter matrix of size $64 \times 576$ with input depth $D_i = 64$ implies a $3 \times 3$ kernel, since $\sqrt{576 / 64} = 3$.
		\end{itemize}
		
		\item \textit{Reconstructing Network Architecture}:
		Sequential analysis of GEMM calls and their constraints enables layer-by-layer recovery, including shortcut connections.
	\end{enumerate}
	
	Cache Telepathy significantly reduces the architecture search space; for instance, the search space for VGG-16 drops from $5.4 \times 10^{12}$ to just 16 candidates, and for ResNet-50 from $6 \times 10^{46}$ to only 512 possibilities.
	
	\subsubsection{DeepCache: Revisiting Cache SCAs on Optimized Deep Neural Network Executables}
	
Another study \cite{7} that targets cache-based side-channel
attacks against optimized executable
files of deep neural networks, aims to infer model architecture remotely in shared cloud settings, even when the attacker has no access to shared libraries or model code. DeepCache addresses the challenges posed by compiler-level optimizations that reduce the effectiveness of traditional cache attacks.
	
	\textbf{Attack Methodology:}
	\begin{enumerate}
		
		\item \textit{Collecting Cache Traces via Prime+Probe}:
		\begin{itemize}
			\item By repeatedly executing Prime+Probe, the attacker collects binary traces representing cache access over time.
		\end{itemize}
		
		\item \textit{Segmenting Traces and Identifying Layer Boundaries}:
		\begin{itemize}
			\item Using Autoencoder + Conv-LSTM (Long Short-Term Memory), the method detects significant shifts in cache patterns, which indicate transitions between layers.
		\end{itemize}
		
		\item \textit{Extracting Features from Segmented Traces}:
		\begin{itemize}
			\item Each trace segment is encoded into a numeric vector via contrastive learning.
			\item Similar layers yield similar feature vectors; dissimilar layers yield distinct ones.
		\end{itemize}
		
		\item \textit{Inferring Layer Types and Parameters}:
		\begin{itemize}
			\item The extracted vectors are compared against a reference database to determine the layer type (e.g., convolution, pooling, fully-connected) and key parameters like kernel size and stride.
		\end{itemize}
	\end{enumerate}
	
\begin{figure}[htbp]
		\centering
		\includegraphics[width=\linewidth]{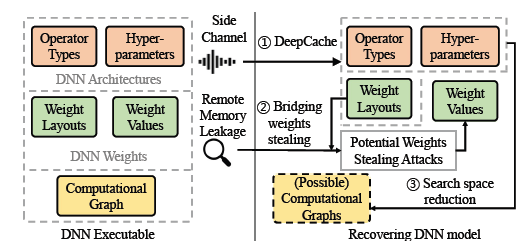} 
		\caption{DeepCache pipeline ~\cite{7}.}
		\label{fig:DeepCacheFig}
	\end{figure}

	DeepCache reveals that compiler optimizations, while improving performance, introduce regular patterns that can still be exploited. These findings highlight a critical vulnerability in how optimized executables are deployed, posing serious risks to model intellectual property. The complete DeepCache leakage–recovery pipeline is shown in Figure~\ref{fig:DeepCacheFig}.

\subsection{Memory Access Pattern-Based Attacks}

Memory access pattern-based attacks analyze how deep learning models interact with main memory such as DRAM to extract sensitive information including model architecture, layer dimensions, and data layout. These attacks do not require direct access to memory contents; rather, they leverage observable temporal or spatial memory access patterns to reverse-engineer the internal structure of the model. Below, we review a representative work that demonstrates the practical feasibility of recovering neural network architectures and parameters via memory behavior.

\subsubsection{Reverse Engineering Convolutional Neural Networks through Side-Channel Information Leakage}
	
	The study in~\cite{9} introduces a novel attack on CNNs deployed on custom accelerators. The main objective is to reconstruct the confidential network architecture and trained parameters (such as weights), even when data remains fully encrypted in memory. The attack utilizes side-channel leakage from memory access behavior in two distinct stages: structure recovery and weight extraction.
	
	\textbf{Attack Methodology:}
	
$\bullet$ \textbf{Reconstructing the Network Structure:}
		\begin{enumerate}
			\item \textit{Identifying Layer Boundaries via RAW Dependency:}
			During CNN execution, the output feature map (OFM) of a layer is first written to memory, then immediately read as the input feature map (IFM) by the next layer. By tracking such write-then-read transitions at the same memory address, the attacker can identify layer boundaries.
			
			\item \textit{Classifying Accessed Data Types:}
			Each CNN layer involves three types of data with distinct memory behaviors:
			\begin{itemize}
				\item Filter weights – read-only, constant throughout.
				\item Input feature maps – read at the start of a layer.
				\item Output feature maps – primarily written, rarely read in the same layer.
			\end{itemize}
			The attacker differentiates these based on observed read/write access patterns.
			
			\item \textit{Estimating Layer Dimensions:}
			These data blocks (IFMs, OFMs, filters) are stored contiguously in memory. By observing memory access granularity and continuity, the attacker estimates their sizes and thereby infers layer-wise dimensional parameters.
			
			\item \textit{Solving Structural Equations:}
			With IFM and OFM dimensions known, the attacker formulates equations describing their relationship with filter size, stride, and padding. Solving these yields precise structural details of each layer.
			
			\item \textit{Refining with Execution Time Analysis:}
			By measuring the runtime of each layer and leveraging the correlation between execution time and the number of MAC operations, the attacker filters out implausible layer configurations to converge on the most accurate structure.
		\end{enumerate}
		
$\bullet$ \textbf{Recovering Weights via Zero Pruning-Based Inference:}
		\begin{enumerate}
			\item \textit{Analyzing Pruned Output Patterns:}
			Many networks skip storing outputs that evaluate to zero to optimize inference. By crafting inputs and tracking transitions from zero to non-zero outputs, the attacker obtains clues about the weight-to-bias ratio.
			
			\item \textit{Locating Zero-Crossing Points via Binary Search:}
			The attacker incrementally perturbs inputs to pinpoint exact thresholds where outputs transition from zero to non-zero. Each transition yields a constraint equation linking weights and biases.
			
			\item \textit{Inferring Weight-to-Bias Ratios:}
			By aggregating multiple such equations, the attacker can deduce the ratio of each weight to its corresponding bias, significantly narrowing the uncertainty about the model parameters.
		\end{enumerate}

	The study demonstrates that using access patterns alone, adversaries can reconstruct networks like AlexNet and SqueezeNet with high fidelity—reducing the architectural search space for AlexNet from an intractable number to just 24 candidates. Moreover, weight-to-bias ratios were inferred with remarkable precision using zero-crossing analysis. These findings highlight a critical threat: even with encrypted data, off-chip memory access patterns can leak sufficient information to compromise the intellectual property and integrity of deep learning models. This underscores the need for countermeasures such as access obfuscation or randomization.

\subsection{Context Switch-Based Attacks on GPUs}

Context switch-based SCAs in GPUs exploit residual information generated during task-switching events in multi-tenant environments to extract sensitive insights about deep learning models. In shared infrastructures such as MLaaS, multiple users may concurrently utilize the same GPU. During context switching, the GPU performs task scheduling and saves/restores execution states, potentially leaking information such as layer execution sequences, input/output dimensions, and operation types~\cite{1,6}. This subsection examines a representative study that empirically demonstrates how such leakages can be exploited to infer a model's architecture.

\subsubsection{Leaky DNN: Stealing Deep Learning Model Secrets via GPU Context-Switch Side Channels}
	
	The study in~\cite{6} introduces a novel SCA designed to extract confidential information from deep neural networks executing on GPUs. The core idea is that in shared environments—such as public clouds—an attacker can utilize context switch-induced latency artifacts to infer detailed structural and hyperparameter-level information, including the types and sequence of layers.
	
	\textbf{Attack Methodology:}
	\begin{enumerate}
		\item \textit{Threat Model}:
		\begin{itemize}
			\item The attacker and victim share the same GPU instance in a cloud environment.
			\item The attacker has no access to the victim’s model or code but can execute custom CUDA kernels on the shared GPU.
		\end{itemize}
		
		\item \textit{Attack Slowdown via Spy Kernels}:
		\begin{itemize}
			\item The attacker executes multiple lightweight spy kernels concurrently to increase the execution time of the victim's layers.
			\item This extended execution window allows for higher-resolution sampling of context-switch latency, improving attack accuracy.
		\end{itemize}
		
		\item \textit{Trace Collection and Analysis}:
		\begin{itemize}
			\item Using the CUPTI profiler, the attacker collects timing traces related to context switch overhead.
			\item These traces encode indirect clues about the victim model's structure and layer specifications.
		\end{itemize}
		
		\item \textit{Layer Classification Using LSTM Models}:
		\begin{itemize}
			\item To identify layer types (e.g., convolutional, fully connected, pooling), the attacker trains LSTM models, well-suited for analyzing sequential time-series data.
			\item The collected latency traces are input to the trained models for classification.
		\end{itemize}
		
		\item \textit{Hyperparameter Inference with Dedicated LSTM Models}:
		\begin{itemize}
			\item Separate LSTM networks are trained to infer specific hyperparameters, including neuron counts, kernel sizes, number of filters, and strides.
		\end{itemize}
		
		\item \textit{Error Correction Using Structural Priors}:
		\begin{itemize}
			\item The attacker applies common architectural design rules (e.g., convolution layers typically precede pooling layers) to refine layer sequences and correct misclassifications.
		\end{itemize}
	\end{enumerate}
	
	The results show that the proposed attack can successfully reconstruct the architectures of well-known models like ZFNet and VGG16 with high precision. These findings underscore a critical vulnerability in multi-tenant GPU environments and highlight the need for effective defensive mechanisms to preserve model intellectual property in such contexts.

\section{Defensive Strategies Against SCAs}
\label{four}

This section reviews major defense mechanisms and detection frameworks proposed to mitigate hardware SCAs targeting deep learning models. For each class of attack—including power-based, EM, and cache-based threats—we discuss specific countermeasures implemented at both hardware and software levels. Additionally, we review a runtime detection technique aimed at identifying cache-based side-channel threats.

To provide a concise overview, Table~\ref{tab:defense-matrix} summarizes the main defense mechanisms discussed in this section. Each method is mapped to the type of SCA it targets, along with the associated performance, power, or area overheads when known. This structured comparison highlights the trade-offs between security and efficiency, and illustrates how certain countermeasures—such as hiding/balanced logic and dummy operations—extend protection beyond power, EM, and cache-based leakages to timing and context-switching channels as well.

\begin{table*}[h]
	\centering
	\caption{Mapping of Defense Mechanisms to Attack Types and Their Overheads}
	\label{tab:defense-matrix}
	\begin{tabular}{|>{\centering\arraybackslash}p{4.5cm}|
			>{\centering\arraybackslash}p{5.5cm}|
			>{\centering\arraybackslash}p{7.cm}|}
		\hline
		\textbf{Defense Mechanism} & \textbf{Targeted Attack Type(s)} & \textbf{Overhead (Area/Power/Latency/Accuracy Impact)} \\ \hline
		
		Masking & Power, EM & Latency $\uparrow$, moderate power overhead\\ \hline
		
		Hiding and Balanced Logic & Power, EM, Timing, Context Switching & Area $\uparrow$, latency $\uparrow$ \\ \hline
		
		Randomized Execution / Shuffling & Timing, Cache-based & Latency $\uparrow$, power $\uparrow$; entropy $\uparrow$ but throughput $\downarrow$ \\ \hline
		
		Noise Injection & Power, EM & Power $ \uparrow$, may reduce efficiency \\ \hline
		
		Physical Shielding / Filtering & Power, EM & Area $\uparrow$, cost $\uparrow$; minimal latency but expensive \\ \hline
		
		Cache Partitioning & Cache-based & Memory overhead $\uparrow$, performance $\downarrow$ in shared systems \\ \hline
		
		Cache Randomization / Line Remapping & Cache-based & Latency $\uparrow$, moderate slowdown \\ \hline
		
		Dummy Operations / Decoy Accesses & Cache-based, Timing, Context Switching & Execution time $ \uparrow$, high computational overhead \\ \hline
		
		Runtime Detection & Cache-based & Moderate latency depending on monitoring intensity \\ \hline
		
	\end{tabular}
\end{table*}
\newpage
\subsection{Countermeasures Against Power and EM Leakage}

Power and EM SCAs exploit physical emissions generated during model execution to infer sensitive information, such as architecture details or model weights. To mitigate such leakages, researchers have proposed various techniques designed to obfuscate the correlation between internal computations and observable physical signals. Key approaches include:

$\bullet$\textbf{Masking:}
	This technique involves randomizing sensitive data—such as weights or intermediate outputs—by combining them with random masks. As a result, power or EM leakage becomes statistically independent from actual data values. These masks are removed after computation to restore correct outputs. Masking is especially feasible for digital arithmetic operations, such as multiply–accumulate units. For example, the approach in~\cite{15} applies masking to protect the internal states of deep learning models. Additionally, \cite{45} presents a masked hardware accelerator for feed-forward NNs with fixed-point arithmetic, protecting against side-channel analysis through hardware-level masking.

$\bullet$ \textbf{Hiding and Balanced Logic:}
	This approach modifies hardware designs so that power consumption and EM emissions become data-independent. In balanced logic implementations, a fixed number of bits switch in each operation, producing uniform power patterns across different inputs. Such uniformity eliminates the attacker’s ability to correlate leakage with specific computations.
	
$\bullet$ \textbf{Randomized Execution and Shuffling:}
	By randomizing the execution order of operations—such as the arrangement of neurons or filters—the power trace associated with each inference becomes unpredictable. This increased entropy disrupts the attacker’s ability to align traces with specific layers or input features, thereby mitigating inference-based extraction \cite{46}.
	
$\bullet$\textbf{Noise Injection:}
	Here, artificial noise is deliberately added to power consumption or EM signals to obscure the original signal. This noise may be random, periodic, or continuous. Although simple and effective in practice, noise injection increases energy consumption. In~\cite{16}, a ring oscillator is used to inject noise and protect model computations.
	
$\bullet$ \textbf{Physical Shielding and Filtering:}
	These methods involve physical modifications to the hardware. For instance, shielding layers (e.g., metal enclosures or absorbing materials) can prevent EM emissions from propagating externally. Similarly, analog filters may be employed to smooth out transient variations in power signals. These techniques are commonly used in military and industrial applications where security requirements are stringent.

\subsection{Countermeasures Against Cache-Based Leakage}

Cache-based SCAs exploit timing behaviors and memory access patterns within multi-level cache hierarchies (e.g., L2, LLC) to infer sensitive model information. Several architectural and system-level defenses have been proposed to address these vulnerabilities:

$\bullet$ \textbf{Cache Partitioning:}
	In this method, cache lines are partitioned and assigned to different processes or virtual machines to prevent cross-core eviction. This isolation ensures that the attacker cannot observe cache usage by the victim, thereby disrupting attacks such as Prime+Probe \cite{47}.
	
$\bullet$\textbf{Cache Randomization and Line Remapping:}
	This approach alters the mapping between memory addresses and cache sets dynamically at runtime. By preventing fixed memory-to-cache correspondence, it becomes significantly harder for attackers to predict or track victim access patterns, thereby neutralizing attacks like Flush+Reload and Prime+Probe \cite{47}.
	
 $\bullet$\textbf{Dummy Operations and Decoy Accesses:}
	These countermeasures insert non-functional memory accesses or spurious instructions into program execution to distort cache traces. For example, random loads/stores, loop repetitions, or dummy function calls can obfuscate the real memory usage of the model, misleading the attacker and increasing their analysis complexity \cite{21}.

While these defense mechanisms can significantly reduce side-channel risks, they often introduce substantial overhead that can negatively impact system performance. Therefore, such countermeasures should be carefully designed and deployed to balance security and efficiency without compromising the overall functionality of the system.

\subsection{Runtime Detection of Cache-Based SCAs}

Beyond architectural countermeasures, runtime detection techniques provide an additional layer of defense by monitoring system behavior during execution. These methods aim to identify anomalous patterns indicative of SCAs and respond in real-time.
For instance, Profiling tools such as \texttt{perf} (a standard Linux utility) are commonly used to monitor low-level hardware events including \texttt{cache-misses}, \texttt{cache-references}, \texttt{branch-mispredictions}, and \texttt{instructions}. These performance counters offer insight into cache-related behavior and can expose attack patterns such as those seen in Flush+Reload or Prime+Probe attacks~\cite{17,19,13}.
A general workflow of run-time detection mechanisms includes:

\begin{itemize}
	\item \textit{Performance Monitoring:} System-level metrics are collected periodically or after specific code blocks using profiling tools like \texttt{perf}.
	\item \textit{Feature Extraction:} Features such as cache miss rates, LLC access frequency, or load/store ratios are extracted to capture cache behavior.
	\item \textit{Machine Learning Classification:} Classifiers such as Support Vector Machines (SVMs), Random Forests, or Neural Networks are trained to distinguish between normal and attack-induced behavior.
	\item \textit{Online Detection and Response:} The trained model is deployed for real-time monitoring. Upon detecting suspicious activity, the system can trigger alerts, log incidents, or terminate compromised processes.
\end{itemize}

Although most existing runtime detection approaches have focused on cryptographic scenarios (e.g., AES or RSA), their principles are readily applicable to machine learning contexts—especially in multi-tenant or cloud environments where model co-location is common. Extending these detection strategies to protect IP in Deep learning models represents a promising avenue for future work.

\section{Future Research Directions}
\label{five}

In recent years, SCAs targeting deep learning hardware have received increasing attention. Nevertheless, several important research gaps remain unaddressed. This section outlines key future research directions that can enhance our understanding of these threats and facilitate the development of more robust defense mechanisms.

$\bullet$	\textbf{Attacks on Emerging Deep Learning Architectures:}
	While most existing studies have focused on CNNs, modern architectures such as Transformers and LSTMs—used widely in natural language processing, time-series analysis, and computer vision—remain largely unexplored in the context of side-channel vulnerabilities. Moreover, scenarios involving Federated Learning and Distributed Learning, where models are trained or deployed across multiple devices, present new challenges for leakage analysis and require deeper investigation.
	
$\bullet$\textbf{Hybrid Hardware–Software Attack Surfaces and Defense Strategies:}
Existing research on SCAs has largely focused on either hardware-based or software-based perspectives in isolation. However, in practical scenarios, attackers often exploit a combination of leakage vectors—such as timing characteristics, cache activity, power usage, and software-level cues (e.g., shared libraries). This highlights the need for integrated frameworks capable of analyzing and mitigating threats that span both hardware and software layers. Advancing such cross-layer defense mechanisms remains a key direction for future investigation.
	
$\bullet$	\textbf{Robust Model Watermarking Techniques:}
	With the increasing adoption of MLaaS platforms, the protection of model IP has become more urgent. Model watermarking—embedding digital signatures within model parameters or designing specific responses to secret trigger inputs—offers a promising means to assert ownership. However, existing watermarking schemes have rarely been evaluated under SCA scenarios. Developing watermarking techniques that are resilient to power, cache, or timing-based leakage would significantly strengthen IP protection in machine learning systems.

\section{Conclusion}
\label{six}

This report has shown that deep learning models remain vulnerable to a broad range of hardware SCAs. These attacks exploit physical indicators such as power consumption, cache access patterns, and timing fluctuations to reconstruct model architectures, extract weights, and even recover sensitive inputs. A review of recent research reveals that such threats are effective not only on CPUs and GPUs but also on custom hardware accelerators. Although a variety of defense strategies have been developed, significant gaps still remain in achieving comprehensive protection. As such, the development of robust and adaptive defense mechanisms at both hardware and software levels is essential to safeguarding the security and intellectual property of deep learning systems.

\bibliographystyle{unsrtnat}
\bibliography{biblio}

\begin{wrapfigure}[8]{l}{1.6cm}
	\begin{center}
		\includegraphics[width=2.4cm, height= 2.6cm]{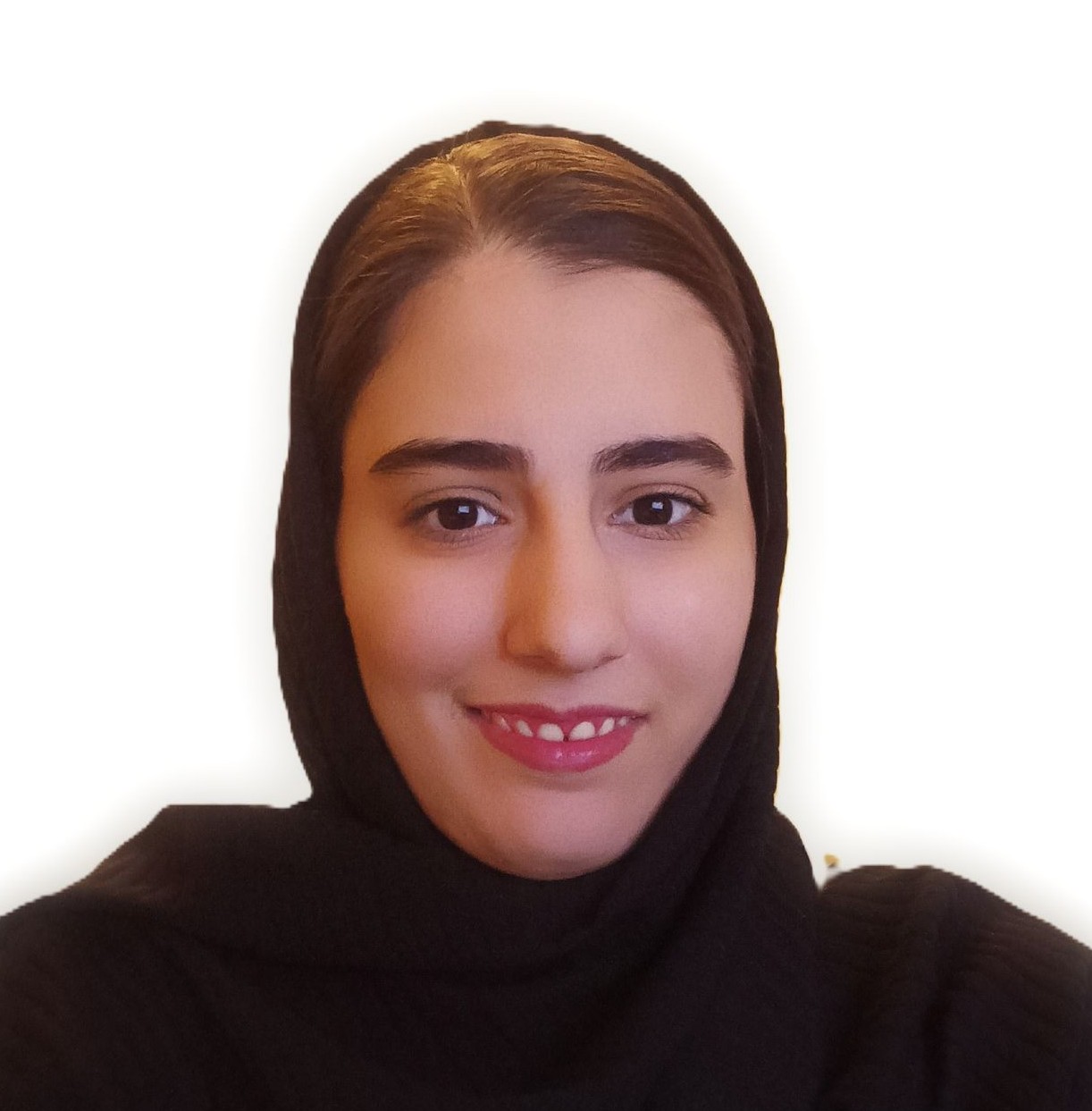}
	\end{center}
\end{wrapfigure}
\noindent
\newline
{\bf Zahra Mohammadi} {is currently a Ph.D. candidate in Computer Engineering at the University of Tehran, specializing in Computer Architecture. Her research primarily focuses on detecting sleep disorders through advanced analysis of biomedical signals, utilizing machine learning and deep learning techniques. In addition, she actively investigates hardware side-channel attacks on deep learning models, with an emphasis on understanding and mitigating hardware-level vulnerabilities. She received her M.S. in Computer Engineering from the University of Tehran in 2023, where she ranked first among all graduates, and her B.S. in Computer Engineering from Shahid Beheshti University in 2020.}

\begin{wrapfigure}[8]{l}{1.6cm}
	\begin{center}
		\includegraphics[trim={7cm 12cm 9cm 13cm}, width=0.9in,height=1.15in,clip]{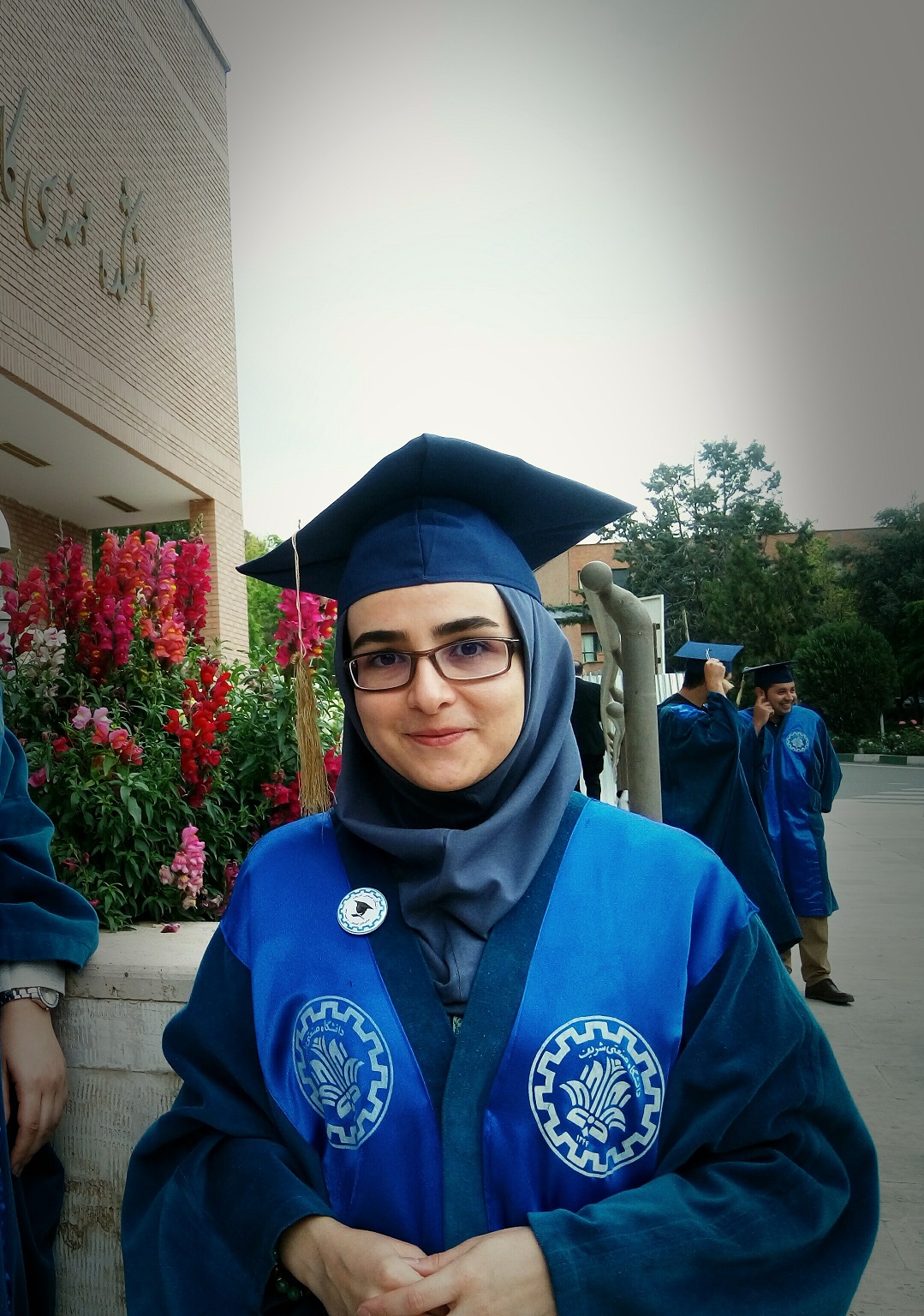}
	\end{center}
\end{wrapfigure}
\noindent
\newline
{\bf Mona Hashemi} {completed her Ph.D. at the University of Tehran, Iran, in 2025. She received her M.S. and B.S. degrees in Computer Engineering from Sharif University of Technology and K. N. Toosi University of Technology in 2016 and 2012, respectively. During her Ph.D., she was a visiting student at the National University of Singapore (NUS), where she is currently a postdoctoral researcher. Her research interests include Very Large Scale Integration Systems (VLSI), Hardware Security and Trust, as well as Efficient and Dependable Computing.}

\begin{wrapfigure}[8]{l}{1.5cm}
	\begin{center}
		\includegraphics[width=2.3cm, height= 2.85cm]{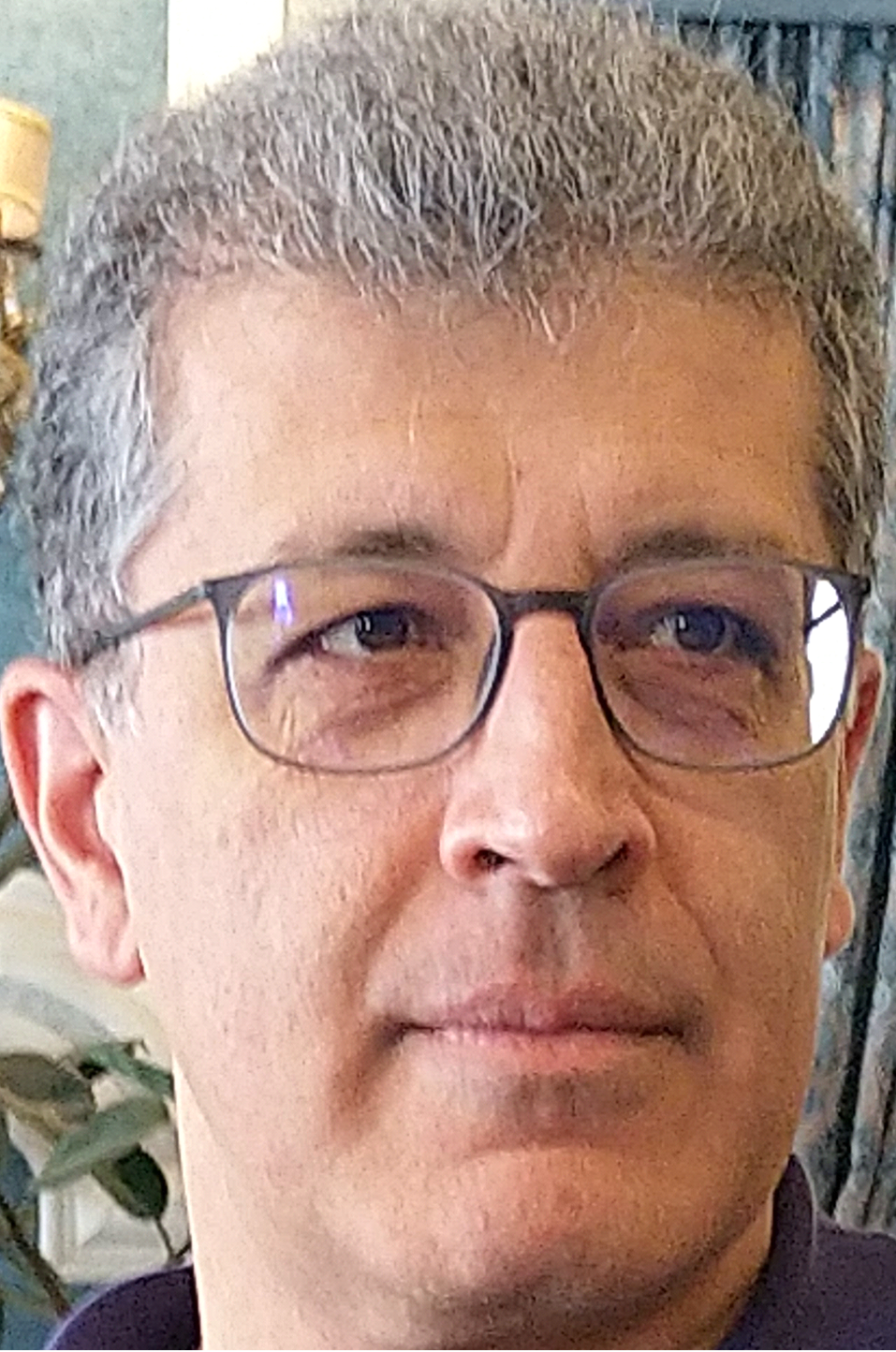}
	\end{center}
\end{wrapfigure}
\noindent
\newline
{\bf Siamak Mohammadi} { received his BSc, MSc and PhD degrees from the University of Paris Sud Orsay, France in 1990, 1992 and 1996, respectively, all in electrical engineering. From 1997 to 1999 he was a Research Associate with the Department of Computer Science, University of Manchester, England. In 1999 he moved to Canada and worked in Semiconductor industry in Toronto until 2005. Currently he is an Associate Professor in School of Electrical and Computer engineering, at the University of Tehran, Iran. He has over 30 years of experience in Low Power Design , Verification,  and Hardware Security of Embedded Systems.}
\end{document}